\documentclass[aps,prl,twocolumn,groupedaddress]{revtex4}
\usepackage{graphicx}

\begin{document}


\title{Inelastic Collisions of a Fermi Gas in the BEC-BCS Crossover}


\author{X.~Du, Y.~Zhang, and J.~E.~Thomas}
\email[jet@phy.duke.edu]{}
\affiliation{Duke University, Department of Physics, Durham, North
Carolina, 27708, USA}



\date{\today}

\begin{abstract}
We report the measurement of inelastic three-body and two-body
collisional decay rates for a two-component Fermi gas of $^6$Li, which are highly
suppressed by the Pauli exclusion principle. Our measurements are
made in the BEC-BCS crossover regime, near the two-body collisional
(Feshbach) resonance. At high temperature (energy) the data shows a
dominant three-body decay process, which is studied as a function of
bias magnetic field. At low energy, the data shows a coexistence of
two-body and three-body decay processes near and below the Feshbach
resonance.  Below resonance, the observed two-body inelastic decay
can arise from molecule-atom and molecule-molecule collisions. We
suggest that at and above resonance, an effective two-body decay
rate arises from collisions between atoms and correlated (Cooper)
pairs that can exist at sufficiently low temperature.
\end{abstract}


\maketitle

Quantum statistics dramatically affects the  inelastic collision
rates that determine the lifetime of cold atomic gases. In an
inelastic three-body collision,  two of the colliding atoms decay to
a bound molecular state, releasing energy.  Interactions between
atoms can be strongly enhanced by tuning a bias magnetic field near
a collisional (Feshbach) resonance~\cite{Feshbachpeople1,
Feshbachpeople2}. In a Bose gas, this enhancement is accompanied by
an inelastic collision rate that increases by two or three orders of
magnitude compared to that obtained away from
resonance~\cite{Roberts2000}, and a correspondingly short lifetime
of just a few ms at typical atomic densities. In contrast, for a
Fermi gas in a mixture of one or two different spin states, the
probability of three atoms colliding is highly suppressed by the
Pauli exclusion principle. The lifetime of the cloud is on the order
of 0.1 s for fermionic $^{40}$K~\cite{Regal2003,Regal2004} and 50 s
for $^6$Li~\cite{Dieckmann2002, Bourdel2004}. The long lifetime of
Fermi gases is essential to the study of strongly interacting Fermi
gases~\cite{O'Hara2002,RMP2008}, which offers unprecedented
opportunities to test nonperturbative theoretical techniques that
apply to exotic systems ranging from high temperature
superconductors to nuclear matter.  Determination of the inelastic
collision rate coefficients in the strongly interacting regime of a
Fermi gas provides new tests of few-body
theories~\cite{Braaten2006,Bedaque2000,Esry1999,Esry2005,Nielsen1999,Petrov2003,Petrov2004,Stoof2008,Helfrich2009}.

In this Letter we report on the precision measurement of three-body
inelastic collision rate constants $K_3$ for an ultracold two-component Fermi gas
in the BEC-BCS crossover regime near a Feshbach resonance. We also
observe two-body inelastic decay below the Feshbach resonance, which
arises from molecules~\cite{Bourdel2004,Petrov2004}. From the data,
we estimate the corresponding rate constants $K_2$. Finally, we
observe two-body decay at and just above the Feshbach
resonance. We suggest that this process arises from correlated
pairs, which is a many-body effect. We load a Fermi gas from a
single beam CO$_2$ laser trap into a CO$_2$ laser standing wave that
is formed by the incoming and retro-reflected beam. The standing
wave produces a potential with a period of 5.3 $\mu$m that is four
times deeper than that of the single beam trap and tightly confining
in the axial direction (along the standing wave). The corresponding
atomic density is up to $10^{14}$/cm$^3$, $\sim 20$ times higher
than that obtained in the single optical trap. This dramatically
increases the inelastic collision rates, making precise  measurement
of the rate constants feasible.

For two-component Fermi gases, three-body inelastic collisions
arise in the BEC-BCS crossover for processes of the form
$F+F+F'\rightarrow F+(FF')$, where $F$ and $F'$ are fermions in
different states and $(FF')$ is a bound molecular state. On
the BEC side of the Feshbach resonance where the scattering length
$a>0$, the three-body decay rate is predicted to scale as
$a^6$~\cite{Petrov2003, Esry2005}, while on the BCS side ($a<0$),
it should scale as $|a|^{2.455}$~\cite{Esry2005}. By contrast, two-body
inelastic collisions can arise from the decay of real molecules,
which exist on the BEC side. These processes take the form either
$(FF')+F\rightarrow(FF')_- + F$ or
$(FF')+(FF')\rightarrow(FF')_-+(FF')$, where $(FF')_-$ is a deeply
bound molecular state. The theory predicts that the decay rate
scales as $a^{-3.33}$ for atom-molecule collisions or $a^{-2.55}$
for molecule-molecule collisions~\cite{Petrov2004}.

In the experiments, a sample of $^6$Li atoms in a 50-50 mixture of
the two lowest hyperfine states  is loaded into a CO$_2$ laser trap
with a bias magnetic field of 840 G, where the two states are
strongly interacting. Evaporative cooling is performed to lower the
temperature of the sample~\cite{O'Hara2002}. The magnetic field is
then changed in 0.8 seconds to  a final magnetic field where we
perform the measurement. Subsequently, the gas is adiabatically
loaded into a CO$_2$ laser standing wave by slowly turning on the
retro-reflected CO$_2$ laser beam. A quasi-two-dimensional Fermi gas
is then formed and absorption images are taken at various times
after the formation of the 2-D system to determine the inelastic decay rate.

At the final optical trap depth, the measured trap oscillation
frequencies in the standing wave are $\omega_{\perp}=2\pi\times3250$
Hz in the transverse directions and $\omega_{z}=2\pi\times83.5$ kHz
in the axial direction. The corresponding frequencies in the single
beam trap are $\omega_{\perp}=2\pi\times1650$ Hz and
$\omega_{z}=2\pi\times56$ Hz, respectively.  Our measurements
indicate very good standing wave alignment, as the transverse
frequency is nearly twice that of the single beam trap, as expected.

 The total energy of the gas obeys the  virial theorem~\cite{ThomasVirial09} when the bias magnetic field is tuned to a broad Feshbach resonance,
 where the Fermi gas is unitary. Since the trap depth is
 large compared to the energy of the cloud, the confining potential $U$ is approximately harmonic. Then
 the total energy is $E=2\langle U\rangle=E_z+E_\perp$, where $E_z$ is the axial energy
 and $E_\perp$ is the transverse energy, referred to the trap minimum.
 We determine only the transverse energy $E_\perp=2m\omega_{\perp}^2{\langle x^2\rangle}$,
 by measuring the mean square transverse cloud size $\langle x^2\rangle$.
 For reference, the transverse energy for the ground state of an ideal two dimensional Fermi
 gas is $E_{I\perp}= \frac{2}{3}E_{F\perp}$, where $E_{F\perp}$ is the transverse Fermi energy,
 $E_{F\perp}=\hbar\omega_{\perp}N_s^{1/2}$. Here $m$ is atomic mass
 of $^6$Li and  $N_s$ is the total atom number in one site.   For our experiments in the unitary gas,
 we measure $E_\perp/E_{F\perp}\sim 1.8$ with $N_s=2,600$ and
 $E_\perp/E_{F\perp}\sim 0.7$ with $N_s=1,600$. If the 2D unitary gas has the same effective mass as the 3D case,
the 2D ground state transverse energy would be $2 E_{F\perp}\sqrt{1+\beta}/3\simeq
0.42\,E_{F\perp}$, using $\beta = -0.60$~\cite{LuoJLTP}. In this case,
our lowest energy would be significantly above the ground state value.

In general, for magnetic fields away from resonance where the
scattering length is finite, the total energy is dependent on the
scattering length~\cite{Werner}. In this case, we measure the
number-independent mean square transverse cloud size $\langle
x^2\rangle/x_{F\perp }^2$, where $x_{F\perp}^{2}$ is defined by
$2m\omega_{\perp}^2x_{F\perp}^2\equiv E_{F\perp}$. For an ideal gas
in the ground state, we note that $\langle
x_0^2\rangle=\frac{2}{3}x_{F\perp}^{2}$.

\begin{figure}[tb]
\includegraphics[width=3.5in]{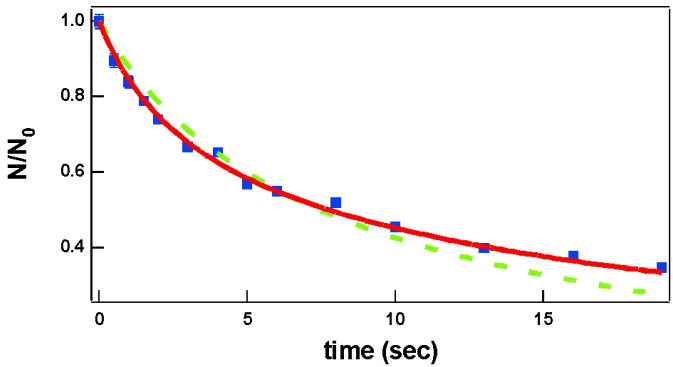}
\caption{Atom number versus time. Data were taken at 834 G and
$E_{\perp}/E_{F\perp}=1.8$. $N$ is total atom number and $N_0$ is initial atom number in the
observed region of the cloud.   Blue dots: Experimental data; Red solid curve:
Three-body decay fit; Green dashed line: Two-body decay
fit.}
 \label{fig:k3at834G}
 \end{figure}
We measure inelastic collision rates by measuring the time
dependence of the atom number and the radial cloud size. The atom
number $N$ as a function of time is~\cite{Roberts2000}
\begin{equation}\label{decay}
    \frac{dN}{dt}=-\Gamma N-\int K_2\, n^2\, d^3x - \int K_3\, n^3\, d^3x,
\end{equation}
where $n$ is the atomic density. On the right side, the first term
arises from background collisions with a density-independent rate
$\Gamma$ ($1/\Gamma=64$ s for our trap). The second term arises
from loss due to two-body inelastic collisions with a rate
coefficient $K_2$, while the third term arises from loss due to
three-body collisions with a rate coefficient $K_3$.

For the conditions of our experiments, where $E_{F\perp}/\hbar\omega_z\simeq 1.5$, the ground axial state contains 90\% of the atoms for an ideal Fermi ga at zero temperature. For simplicity, assume that the 2-D Fermi gas is primarily in the ground axial state of a single site. Then, the atomic density is
\begin{equation}
\label{density}
    n(\rho,z)=\frac{2}{\pi^{3/2}}\frac{N(z)}
    {\sigma_\perp^2\sigma_z}\left(1-\frac{\rho^2}{\sigma_\perp^2}\right)
        \, \exp\left(-\frac{z^2}{\sigma_z^2}\right),
\end{equation}
for $0\leq \rho\leq \sigma_\perp$. Here, $N(z)$ is atom number in the site at position $z$.
$\sigma_\perp$ is transverse width for a fit of a Thomas-Fermi
distribution to the atomic density profile in the transverse
directions, $\sigma_z=(\frac{\hbar}{m\omega_z})^{1/2}$ is axial
width for the ground state (along the standing wave), and
$\omega_z$ is the corresponding axial trap frequency.
\begin{figure}[tb]
\includegraphics[width=3.5in]{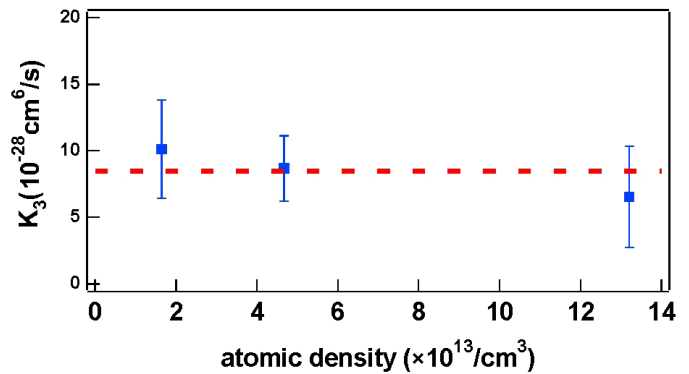}
\caption{Three-body inelastic collision rate coefficient $K_3$
versus atomic density for $E_{\perp}/E_{F\perp}=1.8$. Blue dots: Experimental data. Error bars
indicate statistical errors; Red dashed line: Fit to the data with
$K_3=(8.44\pm1.04)\times10^{-28}$cm$^6$/s.}
 \label{fig:k3vsdensity}
 \end{figure}

In our experiments, $N(z)$ varies as a gaussian distribution
function of $z$ with width $L_z$ over the whole cloud in the axial
direction. Strictly speaking, $\sigma_\perp$, $\sigma_z$ and
$\omega_z$ also vary with $z$ since the depth $U(z)$ of the
potential for a site at $z$  is a Lorentzian function of $z$.
However, we measure a restricted part of the cloud from
$z=-0.83\,L_z$ to $z=0.83\,L_z$ over which $U(z)$ varies less than
10\%. Hence, to good approximation, $\sigma_\perp$, $\sigma_z$ and
$\omega_z$ are spatially constant.

Integrating the atomic density over each well and then over the
restricted region of the cloud, we obtain from Eq.~\ref{decay}
\begin{equation}\label{decay2}
    \frac{dN_c}{dt}=-\Gamma N_c-\alpha_2 K_2   \frac{N_c^2}{\sigma_\perp^2(t)\sigma_z}
     - \alpha_3 K_3  \frac{N_c^3}{\sigma_\perp^4(t)\sigma^2_z},
\end{equation}
where $N_c$ is total number of atoms in the restricted region. Here
$\alpha_2=\frac{2\sqrt{2}}{3} \pi^{-3/2}$ and
$\alpha_3=\frac{2}{\sqrt{3}}\pi^{-3}$. Note that $\sigma_\perp(t)$
is a function of time since heating leads to an increase in
temperature and hence the width of the cloud during the atom loss
process. Typically $\sigma^2_\perp(t)$ and $\sigma^4_\perp(t)$ can
be fit well to exponential curves, $\propto\exp( \gamma t)$.
 Note that at the highest energies used in our experiments,  a significant fraction of atoms can occupy the first axial excited state. If we assume a 50\% fraction, the coefficient $\alpha_3$ is decreased by a factor $0.78$, while $\alpha_2$ is decreased by a factor $0.88$. These systematic corrections are smaller than the statistical uncertainty in our data, so we neglect them in our initial analysis. We then can assume that the axial width is  time independent.

In the first set of experiments, we have measured atom number as a
function of time in the unitary regime at the Feshbach resonance
(834 G), as shown in Fig.~\ref{fig:k3at834G}. The trap depth is set
at 20\% of the maximum attainable by reducing the laser intensity.
The measured transverse energy of the cloud is
$E_\perp/E_{F\perp}=1.8$. We observe a significant ($>60\%$) loss of
the atoms in $\sim$ 20 sec. The data is fit with
Eq.~\ref{decay2}. We find that a three-body decay curve
fits the data very well while a two-body decay curve does not.
This indicates that three-body inelastic collisions play a dominant
role in the atom loss.

Fig.~\ref{fig:k3vsdensity} shows the inelastic decay rate
coefficient $K_3$ as a function of atomic density, at the Feshbach
resonance, for $E_\perp/E_{F\perp}=1.8$. The atomic density is
varied by varying the final trap depth. Data are fit to three-body
decay curves, from which we determine $K_3$. A constant value of
$K_3$ over factor of 10 in atomic density indicates the atom loss
is indeed a three-body decay process. By fitting all of the data
with the same $K_3$, we obtain
$K_3=(8.44\pm1.04)\times10^{-28}$cm$^6$/s.

\begin{figure}[tb]
\includegraphics[width=3.5in]{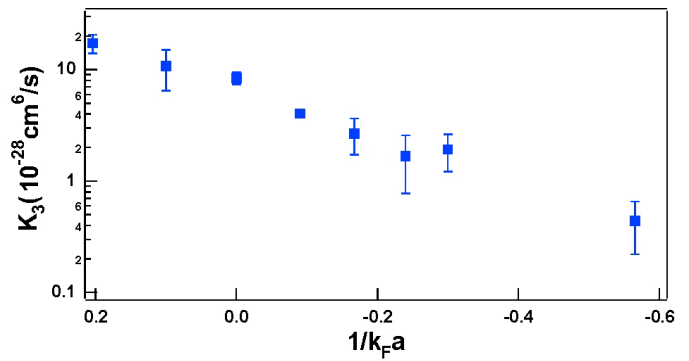}
\caption{$K_3$ versus interaction strength $1/k_{F\perp}a$ at $\langle x^2\rangle /x_{F\perp}^2 =
1.8$. Bars denote statistical error. Varying the magnetic field from 790 G to 1200 G
changes $1/k_{F\perp}a$ from 0.20 to -0.56.
 We observe a factor of 40
change in $K_3$, from $(17.3\pm3.2)\times10^{-28}$cm$^6$/s at 790 G
to $(0.44\pm0.22)\times10^{-28}$cm$^6$/s at 1200 G. }
 \label{fig:k3B}
 \end{figure}

We have also measured $K_3$ as a function of magnetic field
 for $\langle x^2\rangle /x_{F\perp}^2 =
1.8$, which corresponds to the transverse energy
$E_\perp/E_{F\perp}=1.8$ at resonance. The fitted $K_3$ is plotted
as a function of interaction strength $1/k_{F\perp}a$,
Fig.~\ref{fig:k3B}. Here $k_{F\perp}=(2mE_{F\perp})^{1/2}/\hbar$
is the two dimensional Fermi wave vector for an ideal gas at the
trap center and $a$ is the s-wave scattering length. By tuning the
magnetic field from 790 G to 1200 G, we vary $1/k_{F\perp}a$ from
0.20 to -0.56, using the known values of
$a(B)$~\cite{GrimmScattLength}. A factor of $\sim 40$ decrease in
$K_3$ is observed as the bias magnetic field is tuned from the BEC
regime to the BCS regime. We fit our data on the BCS side of the
Feshbach resonance with the function of $K_3=C|a|^n$ and find
$n=0.79\pm0.14$. The result is in significant disagreement with
the theoretical prediction $n=2.455$~\cite{Esry2005}. On the BEC
side, $K_3$ increases as the magnetic field is tuned away from the
Feshbach resonance, instead of peaking on the resonance. This is
consistent with the experiments by other
groups~\cite{Dieckmann2002, Bourdel2004, Regal2004}.

\begin{figure}[tb]
\includegraphics[width=3.5in]{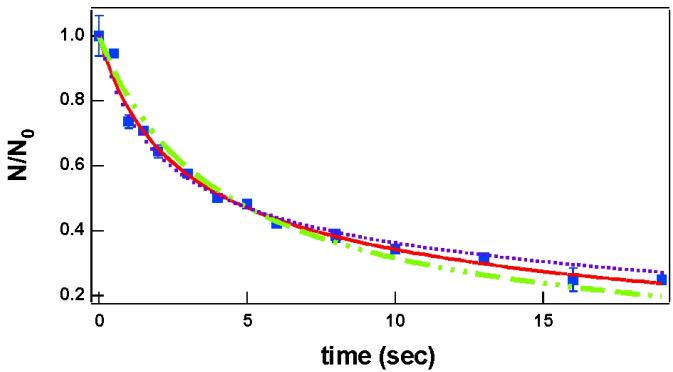}
\caption{Atom number versus time. Data were taken at
$E_{\perp}/E_{F\perp}=0.7$ in the unitary regime. Blue dots: Experimental data; Red solid curve: Combination fit including two-body
and three-body decay; Violet dotted line: Three-body decay
fit; Green dashed line: Two-body decay fit.}
 \label{fig:k2k3}
 \end{figure}

We have repeated the measurement of atom number versus time, at
resonance in the unitary regime, but at a lower energy
$E_\perp/E_{F\perp}=0.7$,  Fig.~\ref{fig:k2k3}. Neither two-body
decay alone nor three-body decay alone fits the data. Instead, the
combination of two-body and three-body decay fits the data well,
which indicates two-body and three-body decays both contribute to
the atom loss.

We suggest that the two-body process is related to correlated
pairs that can exist at low energy (temperature). At higher
energy, only single atoms exist while pairs are broken. In that
case, the Fermi gas can only decay through three-body inelastic
collisions of free atoms. By contrast, at low energy, pair-atom or
pair-pair inelastic collisions are possible. Therefore, both
two-body decay and three-body decay processes can play a role in
the atom loss.

By measuring atom loss as a function of time at
$E_\perp/E_{F\perp}=0.7$, we find
$K_3=(3.30\pm1.81)\times10^{-28}$cm$^6$/s and $K_2=(0.42\pm0.16)\times10^{-14}$cm$^3$/s. It appears that $K_3$
is approximately three times smaller than that at
$E_\perp/E_{F\perp}=1.8$. This suppression cannot arise from Pauli
blocking, as the energetic final states are unoccupied.

The observed scaling of $K_3$ with transverse energy is consistent
with the prediction of Ref.~\cite{Esry2005}, where $K_3\propto E$
for the lowest order process. We observe
$K_3(E_{F\perp}=1.8)/K_3(E_{F\perp}=0.7)=8.44/3.30=2.56$, in very
good agreement with the predicted ratio, $1.8/0.7=2.57$.

Although the data indicates a linear scaling of $K_3$ with energy, a decrease in $K_3$ can also arise from a reduction in the number of
available single atoms, due to pair formation. Defining $f$ as the
fraction of atoms which are paired, the three-body decay rate is
proportional to $(1-f)^3N^3$. For pair-atom collisions, a two-body
rate would scale as $f(1-f)N^2$, while for pair-pair collisions, the
corresponding rate would be proportional to $f^2N^2$.

Using these assumptions, we can rewrite the rate constants that appear in Eq.~\ref{decay2} as
\begin{eqnarray}\label{eq:decay3}
  \nonumber  K_3&\equiv& (1-f)^3 K_{3}^0\\
  K_2&\equiv& f\,K_2^0\equiv f^2K_{2,pp}^0+f(1-f)K_{2,pa}^0.
\end{eqnarray}
Here $K_{2,pa}^{0}$ is the pair-atom
inelastic collision rate coefficient and $K_{2,pp}^{0}$ is the
pair-pair inelastic collision rate coefficient. At
$E_\perp/E_{F\perp}=1.8$, we observe pure three-body decay so that
$f=0$. Hence we have $K_3^0=K_3= (8.44\pm1.04)\times 10^{-28}$
cm$^6$/s.

If we make the extreme assumption that $K_3^0$ is independent of energy, then  we can reinterpret the fitted values of $K_3$ and $K_2$ for
$E_\perp/E_{F\perp}=0.7$ using  Eq.~\ref{eq:decay3} for the rate constants. $K_3=(1-f)^3K_3^0$ yields $f=(30\pm15)\%$ and
$K_2=fK_2^0$ then requires $K_2^0=(1.72\pm1.04)\times10^{-14}$cm$^3$/s. As the fraction of
pairs appears large, it is more likely that the reduction in $K_3$ arises at least in part
from energy scaling, which agrees with
predictions~\cite{Esry2005}, and that the true fraction of pairs is
smaller.

At a magnetic field of 790 G, we first analyze the data to determine $K_3$ and $K_2$ of Eq.~\ref{decay2}.
For $\langle x^2\rangle /x_{F\perp}^2 = 1.8$,  we find $K_3=(17.3\pm3.2)\times10^{-28}$cm$^6$/s. At $\langle x^2\rangle /x_{F\perp}^2 =
0.9$, we obtain, $K_3=(9.35\pm3.06)\times10^{-28}$cm$^6$/s, which is consistent with the predicted linear scaling with energy~\cite{Esry2005}.
The corresponding two-body decay rate constants are $K_2=0$ at $\langle x^2\rangle /x_{F\perp}^2 =
1.8$ and $K_2=(0.57\pm0.22)\times10^{-14}$cm$^3$/s at $\langle x^2\rangle /x_{F\perp}^2 =
0.9$.

If we again assume instead that  $K_3^0$ of Eq.~\ref{eq:decay3} is
independent of energy, we have $K_3^0=(17.3\pm3.2)\times10^{-28}$cm$^6$/s. Using $K_3=(9.35\pm3.06)\times10^{-28}$cm$^6$/s
for $\langle x^2\rangle /x_{F\perp}^2 =0.9$, we require the molecular fraction to be $f=(19\pm7)\%$. Then, we obtain $K_2^0=(3.22\pm0.60)\times10^{-14}$cm$^3$/s.
Note that, on the BEC side, two-body inelastic collisions are expected to be
molecule-atom or molecule-molecule, as predicted~\cite{Petrov2004}.
The increased two-body rate arising from molecules on the BEC side supports
our assumption that  the two-body rate at and just above resonance arises
from correlated pairs. In this case, a many-body theory of inelastic collisions will be needed to replace the few-body theory that is valid  far from resonance.

Above the Feshbach resonance,  we do not observe a two-body decay
process for $1/(k_{F\perp}a)\leq -0.09$, i.e., $B>860$ G. This
suggests that no pairs are formed  for $B>860$ G  at the lowest
energy $E_\perp/E_{F\perp}=0.7$ we achieve.

By comparing the data at high energy and low energy over a wide
range of density, we are able to distinguish between two-body and
three-body processes. This method may provide a probe to determine
the fraction of pairs or molecules in the Fermi gas, once the energy scaling of $K_3$ is fully
established. In the unitary regime, investigation of the energy (or
temperature~\cite{LuoJLTP}) dependence of $K_3$, as well as the
pair fraction, will be an important topic of future work.

This research is supported by the Physics Divisions of the Army
Research Office and the National Science Foundation, and the
Chemical Sciences, Geosciences and Biosciences Division of  the
Office of Basic Energy Sciences, Office of Science, U.S. Department
of Energy. We are indebted to Le Luo and Bason Clancy for help in the initial stages of this work.


 \end{document}